\documentclass[prd,preprint,showpacs,showkeys,preprintnumbers,amsmath,amssymb]{revtex4-1}
\usepackage{amsmath}
\usepackage{amsthm}
\usepackage{amssymb}
\usepackage{amsfonts}
\usepackage{mathrsfs}

\begin{document}

\title{Full Action for an Electromagnetic Field with Electrical and Magnetic 
Charges}

\author{S.~S.~Serova}
 \affiliation{Physics Faculty, St. Petersburg State University, 
   St. Petersburg, Russia}
\author{S.~A.~Serov}
 \email{serov@vniief.ru}
 \affiliation{Russian Federal Nuclear Centre --
   All-Russian Scientific Research Institute of Experimental Physics,
   Institute of Theoretical and Mathematical Physics, Sarov, Russia}

\date{\today}

\begin{abstract}
The paper offers the full action for an electromagnetic field with 
electrical and magnetic charges; 
Feynman laws are formulated for the calculation of the 
interaction cross-sections for electrically and magnetically charged 
particles on the base of offered action within relativistic 
quantum field theory. 
Derived with formulated Feynman rules 
cross-section of the interaction between an elementary particle with magnetic
charge and an elementary particle with electrical charge proves to be
equal zero.
\end{abstract}


\maketitle

\section{Introduction}
\label{sec:introduction}

The papers \cite{dirac1931}, \cite{dirac1948} by Dirac, in which the
existence of magnetic charges was 
related to quantization of electrical charges, were followed by intensively 
searching for magnetic monopoles, however, the search has not yet led to 
success (see, for example, \cite{milton2006}). 
All the attempts of such search were 
complicated by the lack of a satisfactory theoretical description of 
interactions between ordinary electrical and magnetic charges. 
Moreover, Rohrlich in \cite{rohrlich1966} stated the "falsehood of
variational principle for the full 
theory of \dots electrical and magnetic point charges". 
Later, Zwanziger in \cite{zwanziger1971} 
proposed the full action, including an arbitrary constant 4-vector
$n$, with the electromagnetic field propagator being dependent of
arbitrary vector $n$; 
as a result, the cross-section of the interaction of an elementary
particle with magnetic charge and an elementary particle with
electrical charge proves to be dependent of the arbitrary vector $n$ 
(terms in the interaction cross-section, containing arbitrary vector
$n$, can vanish only for interactions of particles with same type of charge). 

For the description of the electromagnetic field with electrical 
and magnetic charges below
we use \textit{two} four-dimensional vector potentials $A_e$ and $B_g$.
Apparently, for the first time two four-dimensional vector potentials
were used for the
description of the electromagnetic field with electrical 
and magnetic charges in the Cabibbo and Ferrari article
\cite{cabibbo1962} -- see, for example, Singleton
review \cite{singleton1996}.
While there is not enough information, we are inclined to suppose, that
quantums of the $A_e$-field and the $B_g$-field are the same.
Another possibility is discussed in works \cite{singleton1995}, 
\cite{ferreira2006}.

In this paper we propose the full action for an electromagnetic field with 
electrical and magnetic charges. 
Particles with both electrical and magnetic charges
are not considered
(compare with the statement by Weinberg in \cite{weinberg1965}, that 
``a magnetic monopole cannot bear a normal charge''). 
Offered full action does not contain arbitrary constant 4-vector $n$.
We formulate Feynman laws for the calculation of the 
interaction cross-sections for electrically and magnetically charged 
particles on the base of offered action within relativistic 
quantum field theory. 
It is shown, that derived with formulated Feynman rules 
cross-section of the interaction between an elementary particle with magnetic
charge and an elementary particle with electrical charge is
equal zero.
Possibly, this explains the failure of the search of magnetic
monopoles with ordinary elementary particle
detectors.

The expressions below are written in Gaussian units; used denotations
are close to that in \cite{landau1988} and \cite{bogoliubov1984}; 
in particular, coordinate indices 
of four-dimensional vectors and tensors are denoted by latin letters
and have values from 0 to 3; a metrics is considered to 
be defined with diagonal metric tensor: $g^{ik}=0$ at $i \ne k$, 
$g^{00}=1$, $g^{11}=g^{22}=g^{33}=-1$; 
the same upper and lower 
coordinate indices of four-dimensional tensors always imply summation; 
if some indices are not coordinate tensor indices, the summation symbol is 
explicitly shown, if necessary; 
the four-dimensional coordinate is 
$x^i  = \left( {ct, {\boldsymbol{r}}}\right)$,
$x_i  = \left( {ct,-{\boldsymbol{r}}}\right)$,
$x^i x_i  = c^2 t^2  - {\boldsymbol{r}}^2 $, 
where $c$ is the light speed constant, $t$ is time, ${\boldsymbol{r}}$
is three-dimensional radius-vector.

\section{Magnetic Monopole in Classic Field Theory}
\label{sec:versus}

It would appear reasonable that the properties of magnetic charges and the 
electromagnetic field generated by them would be similar to the properties 
of electrical charges and the electromagnetic field generated by them: in 
particular, the force lines of the magnetic field, generated by magnetic 
charges, start from/end in the magnetic charges, while the force lines of the 
electrical field generated by currents of magnetic charges are closed, i.e. 
the existence of magnetic charges, in a sense, restores \textit{symmetry} 
between the magnetic and electrical fields. 
Apparently, the equations 
describing the electromagnetic field generated by magnetic charges 
(currents) should be similar to the equations describing the electromagnetic 
field generated by electrical charges (currents). 
At the same time, in view 
of the topological difference between the electromagnetic field generated by 
magnetic charges and the one generated by electrical charges they should be 
described \textit{separately} in equations. 

The electromagnetic field, generated by the current density of electrical 
charges $\left\{ e_a \right\}$ -- $j_e$,-- can be described using the 
antisymmetric four-dimensional tensor of the second rank $F_{A_e }$, 
which can be represented via the four-dimensional vector potential, 
$A_e^i \ {\overset{\mathrm{def}}{=}}\ \left( {\varphi _e,{\boldsymbol{A}}_e} \right)$
in the following form: 
\begin{eqnarray}
F_{A_e }^{ik} 
\ {\overset{\mathrm{def}}{=}}\ 
\frac{{\partial A_{e}^k }}{{\partial x_i }} - 
\frac{{\partial A_{e}^i }}{{\partial x_k }}=
{\partial ^{i} A_{e}^k } - {\partial ^{k} A_{e}^i } .
\label{tensor_F_e}
\end{eqnarray}

The relation between components of tensor $F_{A_e }$ and components of 
(three-dimensional) vectors of electrical field ${\boldsymbol{E}}$ and 
magnetic field ${\boldsymbol{H}}$ is described as
\begin{eqnarray}
E_e^\alpha = - F_{A_e }^{0\alpha } {\quad} 
\left( {\alpha  = 1,2,3}\right); 
{\quad} H_e^1  =  - F_{A_e }^{23} , 
{\quad} H_e^2  =  - F_{A_e }^{31} ,
 {\quad} H_e^3  =  - F_{A_e }^{12} 
\label{E_H}
\end{eqnarray}
or 
\begin{eqnarray}
F_{A_e }^{ik} =
\begin{pmatrix} 
   0     &     -E_e^1 &     -E_e^2 &     -E_e^3\, \\ 
\; E_e^1 &\,\   0     &     -H_e^3 &\:\:\ H_e^2\, \\ 
\; E_e^2 &\:\:\ H_e^3 &\,\   0     &     -H_e^1\, \\ 
\; E_e^3 &     -H_e^2 &\:\:\ H_e^1 &\,\   0\,
\end{pmatrix} \qquad
F_{A_e\, ik} =
\begin{pmatrix} 
\,\,\ 0     &\:\:\ E_e^1 &\:\:\ E_e^2 &\:\:\ E_e^3\, \\ 
\,   -E_e^1 &\,\   0     &     -H_e^3 &\:\:\ H_e^2\, \\ 
\,   -E_e^2 &\:\:\ H_e^3 &\,\   0     &     -H_e^1\, \\ 
\,   -E_e^3 &     -H_e^2 &\:\:\ H_e^1 &\,\   0\,
\end{pmatrix}
.
\label{F_E_H}
\end{eqnarray}

The ordinary Maxwell's equation system, describing an electromagnetic field 
(in vacuum), generated by the current density of electrical charges $j_e$, 
can be written as 
\begin{subequations}\label{std_Maxwell_eqs_e:whole}
\begin{eqnarray}
\frac{{\partial F_{A_e }^{ik} }}{{\partial x_l }} + 
\frac{{\partial F_{A_e }^{kl} }}{{\partial x_i }} + 
\frac{{\partial F_{A_e }^{li} }}{{\partial x_k }} = 0 ,
\label{std_Maxwell_eqs_e:1}
\end{eqnarray}
\begin{eqnarray}
\frac{{\partial F_{A_e }^{ik} }}{{\partial x^k }}
= - \frac{{4\pi }}{c}j_e^i .
\label{std_Maxwell_eqs_e:2}
\end{eqnarray}
\end{subequations}

Since the result of action of operator 
$\frac{{\partial ^2 }}{{\partial x^i \partial x^k }}$ 
(symmetric with respect to indices $i,k$) on 
antisymmetric tensor $F_{A_e }^{ik}$ (similar to any convolution of 
symmetric tensor with the antisymmetric one) identically equals zero, the 
continuity equation describing the charge conservation law follows from 
(\ref{std_Maxwell_eqs_e:2}):
\begin{eqnarray}
\frac{\partial j_e^i}{\partial x^i} =0 .
\label{conserv_j}
\end{eqnarray}

In quantum field theory, equation (\ref{std_Maxwell_eqs_e:2}) is
replaced, according to definition 
(\ref{tensor_F_e}), by the following equation: 
\begin{eqnarray}
-{\partial _{k}}{\partial ^{k}} A_{e}^i 
+{\partial _{k}}{\partial ^{i}} A_{e}^k
\ {\overset{\mathrm{def}}{=}}\ 
\square A_{e}^i + {\partial _{k}}{\partial ^{i}} A_{e}^k
= - \frac{{4\pi }}{c}j_e^i .
\label{std_Maxwell_eqs_2A}
\end{eqnarray}

In the equation (\ref{std_Maxwell_eqs_e:1}) instead of tensor $F_{A_e }$
\textit{pseudo-tensor}
\begin{eqnarray}
F_{A_e }^{\star \, ik} \ {\overset{\mathrm{def}}{=}}\ 
\frac{1}{2!} \, e^{iklm} F_{{A_e}\, lm} ,
\label{dual_F}
\end{eqnarray}
dual to tensor $F_{A_{e}}$, may be used, where $e^{iklm}$ is the
absolutely antisymmetric, unit, four-dimensional 
pseudo-tensor of the fourth rank with weight $W=+1$, which components 
change their signs with permutation of any two indices and 
\begin{eqnarray}
e^{0123} \ {\overset{\mathrm{def}}{=}}\ +1 ,
\label{def_e}
\end{eqnarray}
by definition (pseudo-tensor $e_{iklm} $ has weight $W=-1$ and 
$e_{0123} = -1$). 
With $x$ 
coordinate system replaced by $\bar {x}$ coordinate system (caused by 
changes in the basis of $n$-dimensional vector space), components of 
pseudo-tensor $T$ of weight $W$, which is $r$-times contra-variant and 
$s$-times co-variant, are transformed according to the law (see, for 
example, \cite{schouten1951}): 
\begin{eqnarray}
\bar T _{k'_1 k'_2  \ldots k'_s }^{k_1 k_2  \ldots k_r }  
&\ {\overset{\mathrm{def}}{=}}\ &
\frac{{\partial \bar x ^{k_1 } }}{{\partial x^{i_1 } }}
\frac{{\partial \bar x ^{k_2 } }}{{\partial x^{i_2 } }} \cdots 
\frac{{\partial \bar x ^{k_r } }}{{\partial x^{i_r } }}
\frac{{\partial x^{i'_1 } }}{{\partial \bar x ^{k'_1 } }}
\frac{{\partial x^{i'_2 } }}{{\partial \bar x ^{k'_2 } }} \cdots   
\nonumber\\* &&
\frac{{\partial x^{i'_s } }}{{\partial \bar x ^{k'_s } }}
T_{i'_1 i'_2  \ldots i'_s }^{i_1 i_2  \ldots i_r } 
J^W .
\label{trans_T}
\end{eqnarray}
In contrast to the ordinary tensor's component transformation, there is a 
factor in the form of Jacobian of coordinate transformation
\begin{eqnarray}
J={\frac{{\partial \left( {x^0 ,x^1 , \ldots ,x^{n-1} } \right)}}
{{\partial \left( {\bar x ^0 ,\bar x ^1 , 
\ldots ,\bar x ^{n-1} } \right)}}}
\qquad
\left| J \right| = \sqrt { -\, \mathrm{det}\left( g^{ik} \right)}  = 1
\label{jacobian}
\end{eqnarray}
of power $W$ in equation (\ref{trans_T}). 
The weight of the product of two tensors equals the summarized weight of tensors 
of each factor and convolution of each pair of indices, consisting of the 
identical upper and lower indices, does not change the weight of
tensor; 
all pseudo-tensors considered in the paper have weight $\pm 1$; 
in particular, a 
magnetic charge is a pseudo-scalar, see below, and, hence, we 
are not far from the fore-quoted Weinberg statement. 

The left-hand part of equation (\ref{std_Maxwell_eqs_e:1}) represents
by itself a tensor of the third rank, which is 
antisymmetric in all indices. 
Having lowered indices of this tensor, 
multiplied by $e^{mlik}$ and performed convolution with respect to three 
pairs of identical indices, we obtain equation 
\begin{eqnarray}
\frac{{\partial F_{A_e }^{\star \, ml} }}{{\partial x^l }} = 0 .
\label{std_Maxwell_eqs_e_1dual}
\end{eqnarray}

The relation between components of pseudo-tensor $F_{A_e }^{\star }$, 
which is dual to $F_{A_e }$, and components of (three-dimensional) vectors 
of electrical field ${\boldsymbol{E}}$ and magnetic field
${\boldsymbol{H}}$ is described as 
\begin{eqnarray}
H_e^\alpha = - F_{A_e }^{\star \, 0\alpha } {\quad} 
\left( {\alpha  = 1,2,3}\right); 
{\quad} E_e^1  =  - F_{A_e }^{\star \, 23} , 
{\quad} E_e^2  =  - F_{A_e }^{\star \, 31} ,
 {\quad} E_e^3  =  - F_{A_e }^{\star \, 12} 
\label{E_H_dual}
\end{eqnarray}
or 
\begin{eqnarray}
F_{A_e }^{\star \, ik} =
\begin{pmatrix} 
   0     &     -H_e^1 &     -H_e^2 &     -H_e^3\, \\ 
\; H_e^1 &\,\   0     &     -E_e^3 &\:\:\ E_e^2\, \\ 
\; H_e^2 &\:\:\ E_e^3 &\,\   0     &     -E_e^1\, \\ 
\; H_e^3 &     -E_e^2 &\:\:\ E_e^1 &\,\   0\,
\end{pmatrix} \qquad
F_{A_e\, ik}^{\star } =
\begin{pmatrix} 
\,\,\ 0     &\:\:\ H_e^1 &\:\:\ H_e^2 &\:\:\ H_e^3\, \\ 
\,   -H_e^1 &\,\   0     &     -E_e^3 &\:\:\ E_e^2\, \\ 
\,   -H_e^2 &\:\:\ E_e^3 &\,\   0     &     -E_e^1\, \\ 
\,   -H_e^3 &     -E_e^2 &\:\:\ E_e^1 &\,\   0\,
\end{pmatrix}
,
\label{F_E_H_dual}
\end{eqnarray}
i.e. in pseudo-tensor $F_{A_e }^{\star }$, as compared to tensor 
$F_{A_e }$, components of (three-dimensional) vectors of the electrical 
and magnetic fields change places. 
So, it would appear reasonable, that the 
electromagnetic field generated by the current density of magnetic charges 
$\left\{ {g}_b \right\}$ -- $j_{g}$ could be described using the 
antisymmetric four-dimensional pseudo-tensor of the second rank, 
$F_{B_{g} }$, which is structurally similar to pseudo-tensor 
$F_{A_e }^{\star }$ and described via four-dimensional
pseudovector-potential 
$B_{g}^i  \ {\overset{\mathrm{def}}{=}}\ 
\left( {\psi _{g},{\boldsymbol{B}}_{g}}\right)$ in the following way: 
\begin{eqnarray}
F_{B_{g}}^{ik} 
\ {\overset{\mathrm{def}}{=}}\ 
\frac{{\partial B_{g}^k }}{{\partial x_i }} - 
\frac{{\partial B_{g}^i }}{{\partial x_k }}=
{\partial ^{i} B_{g}^k } - {\partial ^{k} B_{g}^i } .
\label{tensor_F_g}
\end{eqnarray}

Then, instead of equations (\ref{std_Maxwell_eqs_e:whole})-(\ref{std_Maxwell_eqs_2A}), 
(\ref{std_Maxwell_eqs_e_1dual}), to describe the electromagnetic field 
(in vacuum), generated by the current density of magnetic charges
$j_{g}$, one could use the following equations: 
\begin{subequations}\label{std_Maxwell_eqs_g:whole}
\begin{eqnarray}
\frac{{\partial F_{B_{g}}^{ik} }}{{\partial x_l }} + 
\frac{{\partial F_{B_{g}}^{kl} }}{{\partial x_i }} + 
\frac{{\partial F_{B_{g}}^{li} }}{{\partial x_k }} = 0 ,
\label{std_Maxwell_eqs_g:1}
\end{eqnarray}
\begin{eqnarray}
\frac{{\partial F_{B_{g}}^{ik} }}{{\partial x^k }}
= - \frac{{4\pi }}{c}j_{g}^i .
\label{std_Maxwell_eqs_g:2}
\end{eqnarray}
\end{subequations}
\begin{eqnarray}
\frac{\partial j_{g}^i}{\partial x^i} =0 ,
\label{conserv_j_g}
\end{eqnarray}
\begin{eqnarray}
-{\partial _{k}}{\partial ^{k}} B_{g}^i 
+{\partial _{k}}{\partial ^{i}} B_{g}^k
= \square B_{g}^i + {\partial _{k}}{\partial ^{i}} B_{g}^k
= - \frac{{4\pi }}{c}j_{g}^i ,
\label{std_Maxwell_eqs_2A_g}
\end{eqnarray}
\begin{eqnarray}
\frac{{\partial F_{B_{g}}^{\star \, ml} }}{{\partial x_l }} = 0 .
\label{std_Maxwell_eqs_g_1dual}
\end{eqnarray}

To obtain a closed equation system, describing the electromagnetic field 
together with electrical and magnetic charges, equations
(\ref{std_Maxwell_eqs_e:2}) and (\ref{std_Maxwell_eqs_g:2}) must 
be complemented with motion equations for electrical charges 
\begin{eqnarray}
m_a\, c\,\frac{{du_a^i }}{{ds}} = \frac{{e_a }}{c}
\left( {F_{A_e }^{ik}  + F_{B_{g} }^{\star \, ik} } \right) u_{a\, k} 
\label{eq_e}
\end{eqnarray}
and magnetic charges 
\begin{eqnarray}
m_b\, c\,\frac{{du_b^i }}{{ds}} = \frac{{{g}_b }}{c}
\left( {F_{A_e }^{\star \, ik}  + F_{B_{g} }^{ik} } \right) u_{b\, k} .
\label{eq_g}
\end{eqnarray}
In equation (\ref{eq_e}), $m_{a} $ is the $a$-th electrical charge's
mass, $ds = \sqrt {dx^i dx_i} $, 
\begin{eqnarray}
u_a^i \ {\overset{\mathrm{def}}{=}}\ \frac{{dx_a^i }}{{ds}}
\label{def_u}
\end{eqnarray}
is the dimensionless four-dimensional velocity of the $a$-th electrical
charge. 
The same notations are used in equation (\ref{eq_g}). 
In the right-hand parts of equations 
(\ref{eq_e}), (\ref{eq_g}) in parentheses the electrical and magnetic
fields, generated by 
electrical and magnetic charges, are summarized, according to the 
superposition law. 

Equations (\ref{eq_e}), (\ref{eq_g}), 
(\ref{std_Maxwell_eqs_e:2}), (\ref{std_Maxwell_eqs_g:2}) can be
obtained, basing on the 
requirement of 
\begin{eqnarray}
\delta {\mathscr{A}} = 0
\label{delta_A}
\end{eqnarray}
for variation of the full action 
\begin{eqnarray}
{\mathscr{A}} 
\!&=&\! 
- \sum\limits_a \int {m_a cds}
- \frac{1}{c^2 }\int {\left( {A_e^i  + A_{g}^i } \right) j_{{e}\, i}\, d^{\, 4} x}
- \frac{1}{16\pi c} \int {F_{A_e\, ik}\, F_{A_e }^{ik} d^{\, 4} x} 
 \nonumber\\*
&&\!
- \sum\limits_b \int {m_b cds} 
- \frac{1}{c^2 }\int {\left( {B_{g}^i  + B_e^i } \right) j_{{g}\, i}\, d^{\, 4} x}
- \frac{1}{16\pi c} \int {F_{B_{g}\, ik}\, F_{B_{g}}^{ik} d^{\, 4} x} ,
\label{action_A}
\end{eqnarray}
if we vary trajectories of electrical charges, trajectories of magnetic 
charges, vector-potential $A_{e}$ and pseudovector-potential $B_{g}$. 
Integration in first and
fourth terms in the right-hand part of equation (\ref{action_A}) is
performed with respect to the world lines between two given events, 
in the rest terms 
integration is performed with respect to four-dimensional volume; 
variations of functions are assumed to be equal zero on the integration domain 
boundary. 
Instead of the sum of third and sixth
terms in the right-hand part of equation (\ref{action_A}) expression
\begin{eqnarray}
- \frac{1}{16\pi c} \int 
{\left( {F_{A_{e}\, ik}  + F_{B_{g}\, ik}^{\star } } \right)
 \left( {F_{A_{e}}^{ik}  + F_{B_{g}}^{\star \, ik} } \right) d^{\, 4} x} ,
\label{action_A_F2}
\end{eqnarray}
may be used, as variations of products $F_{A_{e}\, ik} F_{B_{g}}^{\star \, ik}$
and $F_{B_{g}\, ik}^{\star } F_{A_{e}}^{ik}$ are equal zero because of 
(\ref{std_Maxwell_eqs_e_1dual}) and (\ref{std_Maxwell_eqs_g_1dual}).
To vary trajectories of electrical charge $e_{a}$, we use the 
expression, describing the current density of this charge via the 
delta-function of three-dimensional argument: 
\begin{eqnarray}
j_{e_a }^i = e_a 
\delta \left( {{\boldsymbol{r}} - {\boldsymbol{r}}_a } \right)
\frac{{dx^i }}{{dt}} ,
\label{def_j_e_a}
\end{eqnarray}
a similar expression for the current density of a magnetic charge 
\begin{eqnarray}
j_{{g}_b }^i = e_b 
\delta \left( {{\boldsymbol{r}} - {\boldsymbol{r}}_b } \right)
\frac{{dx^i }}{{dt}} 
\label{def_j_g_b}
\end{eqnarray}
is used to vary the magnetic charge trajectories; 
\begin{eqnarray}
j_e^i = \sum\limits_a {j_{e_a }^i}, \quad 
j_{g}^i = \sum\limits_b {j_{{g}_b }^i}.
\label{j_e_j_g}
\end{eqnarray}

Because of (\ref{std_Maxwell_eqs_g:2}) density of current $j_{g}$ is a
pseudo-vector and, in consequence of (\ref{def_j_g_b}), the
magnetic charge should be a pseudo-scalar. 
According to (\ref{trans_T}), the sign of pseudo-scalar depends on
the chosen coordinate system (this can be immediate seen from the
expression for the force, acting on magnetic charge in magnetic field)
and, hence, the pseudo-scalar nature of
magnetic charges can be considered as a 
proof of their existence impossibility 
(this issue will be, possibly, discussed later together with 
quantization of charges). 
The pseudo-scalar nature of 
magnetic charges, in a given case, is a direct result of 
transformation 
properties of electrical and magnetic fields under extended Lorentz
group, consisting of Lorentz transformations, spatial rotations 
and inversions (changes of sign of basic vectors).

To take into account the effect on electrical charges of the electromagnetic 
field, generated by magnetic charges, some \textit{effective} 
vector-potential $A_{g}$ (corresponding to the electromagnetic field, 
generated by magnetic charges) is added to the four-dimensional vector 
potential $A_{e}$ in ${\mathscr{A}}$ and a similar term $B_{e}$ is added to the 
pseudovector-potential $B_{g}$. 
Effective potentials $A_{g}$ and $B_{e}$ are described by equations 
\begin{eqnarray}
\partial ^i A_{g}^k  - \partial ^k A_{g}^i 
\ {\overset{\mathrm{def}}{=}}\ 
F_{A_{g} }^{ik} = F_{B_{g}}^{\star \, ik} 
= \frac{1}{2} \, e^{iklm} F_{{B_{g}}\, lm} ,
\label{A_g}
\end{eqnarray}
\begin{eqnarray}
\partial ^i B_e^k  - \partial ^k B_e^i 
\ {\overset{\mathrm{def}}{=}}\ 
F_{B_e }^{ik} = F_{A_e}^{\star \, ik} 
= \frac{1}{2} \, e^{iklm} F_{{A_e}\, lm} .
\label{B_e}
\end{eqnarray}

Using the effective potential, we replace in equations the external 
electromagnetic field, generated by magnetic (electrical) charges, which 
affects an electrical (magnetic) charge, by the same electromagnetic field, 
generated by \textit{fictitious} electrical (magnetic) charges; 
the requirement for such replacement is that magnetic (electrical) 
charges must be absent in the space-time region of interest
[electromagnetic field must not have singularities due to magnetic
(electrical) charges]:
\begin{eqnarray}
j_{{g}}^i \left( x \right) \equiv 0
\label{j_g_0}
\end{eqnarray}
or
\begin{eqnarray}
j_{e}^i \left( x \right) \equiv 0,
\label{j_e_0}
\end{eqnarray}
respectively.
Because we do not consider particles bearing both electrical and 
magnetic charges, this requirement is always met locally, in the close 
vicinity of electrical (magnetic) charge. 
As we use the variational principle 
to derive Lagrange-Euler \textit{differential} equations,
i.e. locally, we may assume, that these requirements are met. 
If requirement (\ref{j_g_0}) is met,
external derivation of antisymmetric four-dimensional tensor of
the second rank $F_{A_{g} }^{ik}$
\begin{eqnarray}
\frac{{\partial F_{A_{g} }^{ik} }}{{\partial x_l }} + 
\frac{{\partial F_{A_{g} }^{kl} }}{{\partial x_i }} + 
\frac{{\partial F_{A_{g} }^{li} }}{{\partial x_k }} ,
\label{dF_A_g}
\end{eqnarray}
written in the form [compare 
with (\ref{std_Maxwell_eqs_e:1}), (\ref{std_Maxwell_eqs_e_1dual})]
\begin{eqnarray}
\frac{{\partial F_{A_{g} }^{\star \, ml} }}{{\partial x^l }} = 
\frac{{\partial F_{B_{g}}^{ml} }}{{\partial x^l }} = 0 ,
\label{dF_A_g_0}
\end{eqnarray}
equals (identically) zero by virtue of (\ref{std_Maxwell_eqs_g:2}) and (\ref{A_g}),
and it is possible to write explicit expression for effective
potential $A_{g}$, using Poincar\'{e} Lemma 
(see, for example, \cite{lang2002} or \cite{kosyakov2007}):
\begin{eqnarray}
A_{g}^k\left( x\right) = 
\int\limits_0^1 {{\zeta }F_{A_{g} }^{ik} \left( {{\zeta }x}
\right)x_i\, d{\zeta }} 
= \frac{1}{2} \int\limits_0^1 {{\zeta }e^{iklm} F_{{B_{g}}\, lm} 
\left( {{\zeta }x}\right)x_i\, d{\zeta }} .
\label{A_g_expl}
\end{eqnarray}
Analogous expression for effective potential $B_e$
\begin{eqnarray}
B_{e}^k\left( x\right) = 
\int\limits_0^1 {{\zeta }F_{B_{e}}^{ik} \left( {{\zeta }x}
\right)x_i\, d{\zeta }} 
= \frac{1}{2} \int\limits_0^1 {{\zeta }e^{iklm} F_{{A_{e}}\, lm} 
\left( {{\zeta }x}\right)x_i\, d{\zeta }} 
\label{B_e_expl}
\end{eqnarray}
may be written, if requirement (\ref{j_e_0}) is met.
We might do not introduce effective potentials $A_{g}$ and $B_e$, but
simply use expressions (\ref{A_g_expl}) and (\ref{B_e_expl}) in the
expression for the full action (\ref{action_A}); i.e. the action
(\ref{action_A}) is ordinary normal action -- compare with the Rohrlich
statement upper.

According to the definition, effective vector-potential $A_{g}^i$
depends on partial derivatives $\partial ^l B_{g}^m$; 
similarly, effective pseodovector-potential $B_e^i$ depends on partial 
derivatives $\partial ^l A_e^m$. 
Consideration of the dependence of $A_{g}^i$ on partial derivatives
$\partial ^l B_{g}^m$ with a varying $B_{g}$ would lead to the
occurrence of a term, linearly depending on the 
value of current density $j_e$, in the \textit{right-hand part} of
equation (\ref{std_Maxwell_eqs_g:2}),
and a similar term, linearly depending on the 
value of current density $j_g$, would occur in the \textit{right-hand part} of 
equation (\ref{std_Maxwell_eqs_e:2}). 
This contradicts our initial assumption that the 
electromagnetic field generated by electrical charges and the 
electromagnetic field generated by magnetic charges are described 
independently and, hence, with varying potentials $A_e$ and $B_{g}$
we should not vary the 
terms of full action, containing the effective potentials. 
In any case, the source-containing terms of equations (\ref{std_Maxwell_eqs_e:2}),
(\ref{std_Maxwell_eqs_g:2}) make no sense to quantization of a free
electromagnetic field in relativistic quantum field theory.

\section{Magnetic Monopole in Relativistic Quantum Field Theory}
\label{sec:qft}

Since a distinction between action (\ref{action_A}) and an ordinary
action without 
magnetic charges is not important from viewpoint of relativistic quantum 
field theory, the Feynman rules for the calculation of the interaction 
cross-sections for elementary particles with electrical or 
magnetic charges can be formulated by generalization of known results of 
relativistic quantum field theory for an electromagnetic field without 
magnetic charges -- see, for example, \cite{bogoliubov1984}, \S~24 or 
\cite{landau1982}, \S~77. 

Operator $S$ ($S$-matrix), which relates amplitudes of the initial
$\Phi \left( -\infty  \right)$ and the final $\Phi \left( \infty  \right)$
states:
\begin{eqnarray}
\Phi \left( \infty  \right)=S\Phi \left( -\infty  \right),
\label{defSmatr}
\end{eqnarray}
-- can be expressed via chronological exponent 
\begin{eqnarray}
S = T\left\{ \exp \left[ {\frac{i}{\hbar} 
\int {{\mathscr{L}}_I \left( x \right)} d^{\, 4} x} \right] \right\},
\label{Tproduct}
\end{eqnarray}
where ${\mathscr{L}}_I \left( x \right)$ is lagrangian of interaction 
\begin{eqnarray}
{\mathscr{L}}_I \left( x \right) = 
- \frac{1}{{c^2 }} {\left( {A_e^k  + A_{g}^k } \right)j_{e\, k}}
- \frac{1}{{c^2 }} {\left( {B_{g}^k  + B_e^k } \right)j_{{g}\, k}}
\label{L_I}
\end{eqnarray}
and 
\begin{eqnarray}
j_{e}^{k} = \sum\limits_a c {e_{a} \bar \psi _a \gamma ^k  \psi _a},\qquad 
j_{g}^{k} = \sum\limits_b c {g_{b} \bar \psi _b \gamma ^k  \psi _b }.
\label{currents_j}
\end{eqnarray}

For propagator of fermions with magnetic charge we have the usual
expression in terms of causal function $D^c\left( {x - y} \right)$ for scalar
particle of mass $m_{b}$ -- compare with \cite{bogoliubov1984}, (15.17), (22.9):
\begin{eqnarray}
\left\langle {T\left\{\psi _{b} \left( x \right)\bar 
\psi _{b} \left( y \right)\right\}} \right\rangle  = 
- i \left( {i{\gamma }^k {\partial }_k+ \frac{m_{b} c}{\hbar}} \right)
D^c\left( {x - y} \right) .
\label{p_prop}
\end{eqnarray}

Motion equations for free electromagnetic field [compare 
with (\ref{std_Maxwell_eqs_2A}), (\ref{std_Maxwell_eqs_2A_g})]
\begin{eqnarray}
\square A_{e}^k + {\partial ^{k}}{\partial _{l}} A_{e}^l = 0
\label{std_Maxwell_eqs_2A_f}
\end{eqnarray}
and
\begin{eqnarray}
\square B_{g}^k + {\partial ^{k}}{\partial _{l}} B_{g}^l = 0 
\label{std_Maxwell_eqs_2A_g_f}
\end{eqnarray}
have the same form, that in the usual electrodynamics.
Thus, the usual expression for chronological coupling of operators of
electromagnetic field, generated by electrical
charges, may be derived (in the Feynman gauge)
\begin{eqnarray}
\left\langle {T\left\{A^k_{e}\left( x \right)A^l_{e} \left( y \right)\right\}}
\right\rangle
= i{4\pi }g^{kl} D_0^c \left( {x - y} \right) ,
\label{A_e_prop}
\end{eqnarray}
and analogical expression may be derived for chronological coupling of
operators of electromagnetic field, generated by magnetic charges,
\begin{eqnarray}
\left\langle {T\left\{B_{g}^k \left( x \right)B_{g}^l \left( y \right)\right\}}
\right\rangle
= i{4\pi }g^{kl} D_0^c \left( {x - y} \right) .
\label{B_g_prop}
\end{eqnarray}

Because only vector potential $A_{e}$ enters in the equation
(\ref{std_Maxwell_eqs_2A_f}), and only pseudovector-potential
$B_{g}$ enters in the equation (\ref{std_Maxwell_eqs_2A_g_f}),
the chronological coupling of operators of these potentials,
expressed via Green function of motion equations, is equal zero
\begin{eqnarray}
\left\langle {T\left\{A^k_{e}\left( x \right)B^l_{g} \left( y \right)\right\}}
\right\rangle = 0 .
\label{A_e_B_g_prop}
\end{eqnarray}
Hence, as accordingly to (\ref{A_g}), (\ref{B_e})
the effective potential $A_{g}$ is a linear function of $B_{g}$, 
and the effective potential $B_{e}$ is a linear function of $A_{e}$, 
chronological couplings
\begin{eqnarray}
\left\langle {T\left\{A^k_{e}\left( x \right)A^l_{g} \left( y \right)\right\}}
\right\rangle = 0 ,
\label{A_e_A_g_prop}
\end{eqnarray}
\begin{eqnarray}
\left\langle {T\left\{A^k_{g}\left( x \right)B^l_{e} \left( y \right)\right\}}
\right\rangle = 0 ,
\label{A_g_B_e_prop}
\end{eqnarray}
\begin{eqnarray}
\left\langle {T\left\{B^k_{e}\left( x \right)B^l_{g} \left( y \right)\right\}}
\right\rangle = 0
\label{B_e_B_g_prop}
\end{eqnarray}
are also equal zero.

Using expression (\ref{A_g_expl}) for $A_{g}\left( x\right)$,
we obtain for the chronological coupling of
operators $A_{g}\left( 0\right) $ and $B_{g}\left( y\right) $:
\begin{eqnarray}
\!\!\!\!\!\!
\left.{\left\langle {T\left\{A^k_{g}\left( x \right)
B^q_{g} \left( y \right)\right\}}
\right\rangle}\right|_{x = 0}
=
\left.{\left\langle {T\left\{
{\left[ e^{klmn} x_l \int\limits_0^1 {{\zeta } 
{\frac{{\partial B_{g\, m}\left( {{\zeta }x}\right) }}
{{\partial x^n }}}\, d{\zeta }}\right]}
B^q_{g} \left( y \right)\right\}}
\right\rangle}\right|_{x = 0}=0 .
\label{A_g_B_g_prop0}
\end{eqnarray}
By virtue of the homogeneity of the space-time and the arbitrariness
of $y$ in (\ref{A_g_B_g_prop0}) we can conclude, that
\begin{eqnarray}
\left\langle {T\left\{A^k_{g}\left( x \right)
B^q_{g} \left( y \right)\right\}}\right\rangle \equiv 0 .
\label{A_g_B_g_prop}
\end{eqnarray}
Similarly, for electromagnetic field, generated
by electrical charges,
\begin{eqnarray}
\left\langle {T\left\{A^k_{e}\left( x \right)
B^q_{e} \left( y \right)\right\}}\right\rangle \equiv 0 .
\label{A_e_B_e_prop}
\end{eqnarray}

Element of the $S$-matrix, corresponding to the interaction of
elementary particle with magnetic charge and elementary particle with
electrical charge, is a sum of terms, proportional to the
chronological coupling (\ref{A_g_B_g_prop}) or (\ref{A_e_B_e_prop}),
and so it proves to be equal zero; 
therefore, interaction
cross-section of an elementary particle with magnetic
charge and an elementary particle with electrical charge is equal zero.
For the magnetic monopoles search
one may try to use detectors of weak magnetic field instead of
ordinary detectors of elementary particles, based on the interaction
of moving (electrically) charged particles with detector's electrical charges.

\begin{acknowledgements}
The authors would like to thank Prof. D.~Singleton and
Prof. B.~P.~Kosyakov for useful considerations
of some questions, discussed in the article.
\end{acknowledgements}


\begin{thebibliography}{1}

\bibitem{dirac1931}P.~A.~M.~Dirac, 
Proc. R. Soc. Lond., \textbf{A133}(821), 60-72 (1931)

\bibitem{dirac1948}P.~A.~M.~Dirac, 
Sci. Am., \textbf{208}(5), 45-53 (1963)

\bibitem{milton2006}K.~A.~Milton, 
Rep. Prog. Phys., \textbf{69}(6), 1637-1711 (2006)

\bibitem{rohrlich1966}F.~Rohrlich, 
Phys. Rev., \textbf{150}, 1104-1111 (1966)

\bibitem{zwanziger1971}D.~Zwanziger, 
Phys. Rev., \textbf{D3}, 880-891 (1971)

\bibitem{cabibbo1962}N.~Cabibbo and E.~Ferrari, 
Nuovo Cimento, \textbf{23}, 1147-1154 (1962)

\bibitem{singleton1996}D.~Singleton, 
Am.~J.~Phys., \textbf{64}, 452-458 (1996)

\bibitem{singleton1995}D.~Singleton, 
Int.~J.~Theor.~Phys., \textbf{34}, 37-46 (1995);
D.~Singleton, 
Int.~J.~Theor.~Phys., \textbf{35}, 2419-2426 (1996)

\bibitem{ferreira2006}P.~Castelo Ferreira, 
J. Math. Phys., \textbf{47}, 072902 (2006);
P.~Castelo Ferreira,
Europhys. Lett., \textbf{79}, 20004 (2007)

\bibitem{weinberg1965}S.~Weinberg, 
Phys. Rev., \textbf{B138}, 988-1002 (1965)

\bibitem{landau1988}L.~D.~Landau, E.~M.~Lifshitz:
{The Classical Theory of Fields}, Fourth Edition: Volume 2 
(Course of Theoretical Physics Series).
Pergamon Press, Oxford (2000)

\bibitem{bogoliubov1984}N.~N.~Bogoliubov, D.~V.~Shirkov:
{Introduction to the Theory of Quantized Fields}. 
Wiley, New York (1980)

\bibitem{schouten1951}J.~A.~Schouten: 
{Tensor analysis for physicists}. 
Clarendon Press, Oxford (1951)

\bibitem{lang2002}S.~Lang: 
{Introduction to Differentiable Manifolds}. 
Springer, New York (2002)

\bibitem{kosyakov2007}B.~P.~Kosyakov: 
{Introduction to the Classical Theory of Particles and Fields}. 
Springer, Berlin (2007)

\bibitem{landau1982}V.~B.~Berestetskii, E.~M.~Lifshitz, L.~P.~Pitaevskii:
{Quantum Electrodynamics}, 2nd Edition 
(L.~D.~Landau, E.~M.~Lifshitz, Course of Theoretical Physics, Volume 4).
Butterworth-Heinemann, Oxford (1982)

\end{thebibliography}
\end{document}